# Enhanced Traffic Congestion Management with Fog Computing: A Simulation-based Investigation using iFog-Simulator

Alzahraa Elsayed[1], Khalil Mohamed[2], Hany Harb[3]

[1,2,3]Systems and Computers Engineering Dept., Faculty of Engineering, Al-Azhar University, Nasr city 11765, Cairo, Egypt.

*Abstract:* Accurate latency computation is essential for the Internet of Things (IoT) since the connected devices generate a vast amount of data that is processed on cloud infrastructure. However, the cloud is not an optimal solution. To overcome this issue, fog computing is used to enable processing at the edge while still allowing communication with the cloud. Many applications rely on fog computing, including traffic management. In this paper, an Intelligent Traffic Congestion Mitigation System (ITCMS) is proposed to address traffic congestion in heavily populated smart cities. The proposed system is implemented using fog computing and tested in a crowded city. Its performance is evaluated based on multiple metrics, such as traffic efficiency, energy savings, reduced latency, average traffic flow rate, and waiting time. The obtained results are compared with similar techniques that tackle the same issue. The results obtained indicate that the execution time of the simulation is 4,538 seconds, and the delay in the application loop is 49.67 seconds. The paper addresses various issues, including CPU usage, heap memory usage, throughput, and the total average delay, which are essential for evaluating the performance of the ITCMS. Our system model is also compared with other models to assess its performance. A comparison is made using two parameters, namely throughput and the total average delay, between the ITCMS, IOV (Internet of Vehicle), and STL (Seasonal-Trend Decomposition Procedure based on LOESS). Consequently, the results confirm that the proposed system outperforms the others in terms of higher accuracy, lower latency, and improved traffic efficiency.

*Index Terms:* Fog Computing; Traffic Congestion; Traffic Index; Cloud Computing; Traffic Sensing Systems; iFog-Simulator.

## I. INTRODUCTION

The number of connected devices globally is projected to reach 55.9 billion by 2025, with 75% of them connected to IoT platforms, according to a report by the International Data Corporation (IDC). IDC also predicts that the data generated by IoT devices will increase from 13.6 ZB in 2019 to 79.4 ZB by 2025. Smart cities rely on real-time or contextual data analysis to achieve consistent results through collaboration across various sectors. However, the growth of motor vehicles in urban areas has created traffic pressures and issues, leading to the development of traffic signal control using Information and Communication Technique (ICT) [1].

Cisco has stated that current cloud models are inadequate for handling the velocity, diversity, and volume of data generated by the IoT [2]. The growth of urban population, urbanization, and IoT applications presents significant challenges, requiring the use of the latest technological advancements to make cities smarter.

Existing control strategies must overcome long decision and response latency from data processing. The challenge for centralized computation infrastructure is to instantaneously respond to adaptive signal control systems.

Cloud computing environments provide storage and processing power to alleviate the burden on local systems [3].

However, relying solely on the cloud computing model has some drawbacks. For instance, data transmission to cloud servers may require wider network bandwidth, and the service latency may increase as well [4, 5].

These drawbacks can be critical issues for applications such as security systems. To address these problems, researchers have recommended two models: fog computing and edge computing [6].

In several IoT applications, analysis must be conducted instantly, and preventive action must be taken immediately. This means that the opportunity to prevent potential damage decreases during data transmission from the device to the cloud for analysis. Cisco has published a report on this challenging issue that disrupts IoT systems [7]. As the report indicates, IoT systems require a new computing model to deal with the variety, rapidity, and quantity of IoT data. Therefore, instead of using centralized clouds, fog computing uses decentralized resources at the network's edges to process data that is closer to the users [8, 9].

Fog computing, a novel and emerging model, aims to minimize latency, conserve network bandwidth, address security concerns, improve reliability, and ensure secure data collection across a large geographic area with varying environmental conditions. It also involves moving data to the optimal location for processing, distinguishing it from edge computing, which refers to communication, processing power, and storage in end devices [5, 10].

Fog computing is considered a superior security model to cloud computing as it enables local data analysis and offers the benefit of saving network bandwidth, resulting in reduced operational costs. Additionally, the fog computing model excels in making rapid decisions and minimizing arrival time. In contrast, cloud computing often experiences longer data transfer times, leading to increased latency and a higher risk of data loss. Consequently, the fog computing model emerges as a more effective solution compared to cloud computing.

As shown in Fig. 1, fog computing acts as the middle layer between devices and the cloud. The fog computing model for the Internet of Things (IoT) operates at the network edge to bring processing and networking closer to users and data sources, thereby reducing latency and improving processing and communication efficiency [11, 12].

This model is applicable in various applications, such as IoT, mobile applications, geographically distributed applications, delay-sensitive real-time applications, crossroads applications, and greenhouse applications. A fog node can be any device that has computing, storage, and network connectivity [13, 14].

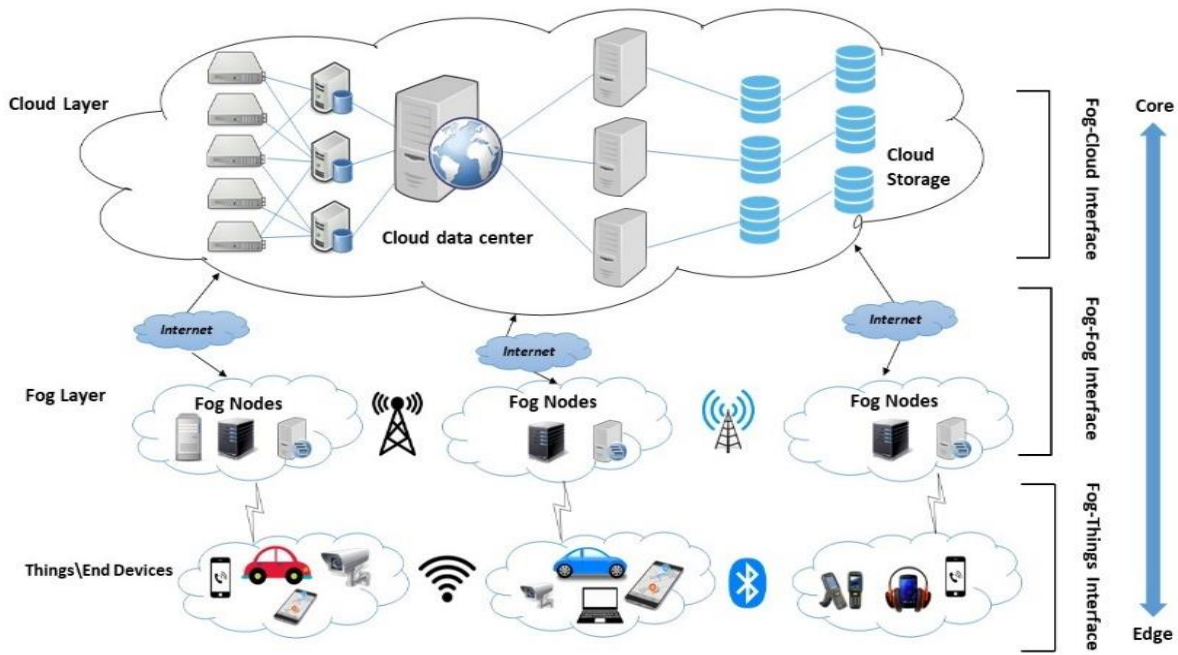

Fig. 1 An IoT Fog Computing Architecture

Smart traffic lights are an example of how integrating information and communication technologies can improve the quality and efficiency of public services. Real-time transportation data is uploaded to the fog to provide transport officials and consumers with updates on the city's transportation availability and conditions [15, 16].

As the urban population grows, effective solutions are required to overcome transportation problems and reduce traffic congestion. To reduce congestion in high-traffic cities, an open fog computing model-based traffic control system is preferable to cloud computing [16]. This is because with fog computing, data is analyzed locally, making it a more secure and cost-effective option that saves network bandwidth. Additionally, fog computing can make decisions more quickly, reducing arrival time. Conversely, cloud computing takes longer to transfer data, increasing latency and the risk of data loss.

This paper introduces an Intelligent Traffic Congestion Mitigation System (ITCMS) that utilizes fog computing to address the issue of traffic congestion in densely populated smart cities. Fog computing extends the capabilities of cloud computing by processing data at the network edge, reducing communication bandwidth and enabling real-time and Internet of Things (IoT) applications. It offers several advantages, such as high processing speed, enhanced security and privacy, and location awareness. To ensure reliable connectivity, collaboration between IoT devices, platforms, and network providers is crucial in establishing stable connection points within the fog network.

The ITCMS is installed on traffic lights and comprises a camera and three LEDs. The camera detects crowded and non-crowded roads, enabling informed decisions regarding vehicle movement. The proposed system is implemented within an environment consisting of four roads (x, y, z, and w) where simultaneous presence of multiple cars leads to traffic congestion. Each fog node is equipped with one camera and three LEDs (Yellow, Green, and Red). As a car travels from R1 (source) to R2 (destination), the LED in the fog node corresponding to R2 turns green, while the LEDs in the other roads turn red.

**The main contribution of this paper is:**
- The proposed solution for mitigating traffic congestion in densely populated urban areas involves the implementation of the Intelligent Traffic Congestion Mitigation System (ITCMS).
- The proposed ITCMS has been evaluated for its effectiveness, and the results have been analyzed to demonstrate its efficiency. The findings suggest that the ITCMS can significantly enhance traffic efficiency, conserve energy, and reduce latency, the average traffic flow rate, and waiting time.
- The system performance of the proposed system is evaluated based on four critical parameters: CPU usage, heap memory usage, throughput, and the total average delay. These parameters are essential in determining the system's performance. Moreover, a comparison has been drawn between the ITCMS and the recently implemented systems for solving the same problem, namely IOV and STL, using two parameters: throughput and the total average delay. This comparison provides valuable insights into the strengths and weaknesses of each system.

The remainder of the paper is structured as follows: Section II discusses related works, while Section III presents the mathematical framework of the proposed (ITCMS). Section IV outlines the structure of the ITCMS, while Section V details its implementation. Lastly, in Section VI, the simulation results are discussed, and the paper is concluded and summarized in Section VII.

II. RELATED WORKS

Given the significant impact of traffic congestion on people's daily lives, it is imperative to address this issue effectively. Numerous efforts have been made to develop traffic management strategies, specifically focusing on the creation of smart traffic signal systems aimed at alleviating traffic congestion. In [17], a system is proposed that uses intelligent traffic lights in fog computing to reduce and manage traffic congestion. This computer platform employs a Q-learning algorithm to control real-time traffic flow by leveraging the capabilities of fog computing for data processing.

Based on categorization and forecasting approaches, the authors of [18] offer a multi-agent system (MAS) strategy for intelligent urban traffic management. They recommend using the k-nearest neighbor and random tree classification models to forecast traffic flow with the greatest accuracy.

The suggested system in [19] combines video monitoring and a big data analytics algorithm to regulate traffic. It introduces a traffic event detection framework.

Gupta et al. [20] proposed a system that connects vehicles and traffic lights directly via the Internet of Things (IoT), where vehicles provide real-time information to the traffic controller regarding road transportation. However, their system is impractical due to its high cost, as each vehicle must have an IoT device.

Rathore et al. [21] proposed a system model that monitors traffic congestion in real-time and tracks any driving violations. The proposed system relies on a camera installed on the back of the vehicle to monitor all preceding vehicles. The fog device connected to the car's camera detects the violation and sends it to the authorities. However, their system is also impractical and costly.

Cunha et al. [22] employed wireless technology to design a traffic management system, creating a model that uses XBEE and a microcontroller connected to several wireless sensors. In [23], a traffic control system model utilizing cloud computing is proposed. Perumalla et al. [24] developed a model that uses an IoT board, microcontroller board, and GPS module. In [25], a system that uses IoT to collect and process traffic data is presented.

The implementation of fog computing in a wide range of practical applications is gaining momentum. Badiea et al. [26] propose a fog computing-based pseudonym authentication (FC-PA) scheme for 5G-enabled vehicular networks. The scheme provides support for batch signature verification, privacy preservation, and pseudonym authentication. It uses a single scalar multiplication operation of elliptic curve cryptography for data verification. The FC-PA scheme is designed to be secure under the random oracle model and is resilient against common security attacks.

In [27], Zeyad et al. propose a fog computing technique for 5G-enabled vehicle networks that utilizes the Chebyshev polynomial. This scheme enables the revocation of pseudonyms and employs fog computing to generate security parameters and validate the authenticity of vehicles.

An efficient mutual authentication scheme is proposed in [28] for 5G-enabled vehicular fog computing, specifically in the context of COVID-19. The scheme has two distinct aspects that rely on a designated flag value to differentiate between regular vehicles and those associated with COVID-19.

Badiea et al. present ANAA-Fog, an anonymous authentication scheme designed specifically for 5G-enabled vehicular fog computing in [29]. This scheme aims to address privacy and security concerns related to real-time services in the context of smart transportation. ANAA-Fog leverages a fog server to generate temporary secret keys for vehicles participating in the system, thereby ensuring the authentication and integrity of exchanged messages.

Zeyad et al. present an efficient authentication scheme for 5G-enabled vehicular networks that leverages fog computing in [30]. The scheme is designed to tackle privacy and security concerns while simultaneously minimizing communication and processing costs. A key component of the scheme is the introduction of a fog server, which is responsible for establishing public anonymity identities and generating corresponding signature keys for authenticated vehicles.

However, most works focus on resource sharing among vehicles, with relatively little emphasis on traffic signal control. Existing traffic control strategies that leverage fog computing primarily aim to improve safety and efficiency, with limited attention given to traffic light optimization. However, fog computing has promising potential for optimizing traffic phase timing due to its low response latency, location awareness, and geographic distribution capabilities. Future studies will focus on designing a scheme with better scalability and compatibility. This scheme incurs slightly higher communication and computing costs than similar studies, but it demonstrates effectiveness in achieving privacy and security objectives.

In this paper, we propose an architecture for traffic signal optimization based on fog computing, which can be readily implemented in real-world scenarios.

## III. A MATHEMATICAL FRAMEWORK OF ITCMS

In this section, we will discuss the mathematical calculations used to compute the green time in traffic signal for each road. We employ light equations in a real-time application to minimize processing time. Eq. (1) is used to calculate the total number of vehicles, while Eq. (2) is used to calculate the time for a single cycle. In Eq. (2), $N_{tc}$ is the total number of vehicles across all roads, $N_c$ is the number of vehicles on each road, and $T_t$ is the total time required to complete a single cycle of traffic lights. The time required for each vehicle to cross the traffic light, μ, is 2.5 seconds in the simulation results.

$$N_{tc} = \sum_{c=1}^{8} N_c \qquad (1)$$

$$T_t = \mu * N_{tc} \qquad (2)$$

$$k = \frac{N_c}{N_{tc}} \qquad (3)$$

$$T_r = k * T_t \qquad (4)$$

To determine the green time in traffic signal on a specific road, we need to make the traffic light more reliable using Compensation Equations (3) and (4). In Equation (3), k represents ratio between the total number of vehicles inside the roads and the number of vehicles on a particular road. In Equation (4), Tr represents the green time at a particular road.

A comparison between previous works was concluded in Table 1.

Table 1: Recent developments in traffic control systems.

| Techniques | Year | Methodology | Advantages | The drawbacks |
|---|---|---|---|---|
| Using fog Computing platform in data control of real-time traffic flow in Intelligent traffic light [17]. | 2021 | To propose an Intelligent traffic light (ITL) control technique, use the Q learning algorithm with fog computing. | High performance & Efficiency. | Priority was not given to emergency vehicles. |
| Data from wireless sensor networks are used in a multi-agent system for intelligent urban traffic control. [18]. | 2022 | Forecasts traffic flow with highest accuracy using k-nearest neighbor and random tree classification algorithms. | Improving inter-agent communication. | Cannot handle circumstances such as ambulances going by, VIP visits, and other catastrophes. |
| Efficient Big Data Analytics-Based Traffic Management Strategies [19]. | 2022 | The framework traffic event detection system is introduced. | Big data should be managed efficiently. | There is no adaptive traffic signal management. |
| Smart Traffic Light System Based on IoT for Smart Cities [20]. | 2021 | A smart system that connects vehicles and traffic lights via cloud computing is proposed. | Minimize the amount of complexity. | Lane changes are not considered. |
| Smart traffic control: Identifying driving-violations using fog devices with vehicular cameras in smart cities [21]. | 2021 | proposed a system model that monitors traffic congestion in real-time and tracks any driving violations. | Detect the violation and sends it to the authorities. | Their system is also impractical and costly |
| Traffic Lights Control Prototype Using Wireless Technologies [22]. | 2016 | Employed wireless technology to design a traffic management system. | Uses XBEE and a microcontroller connected to several wireless sensors. | The model is complex and expensive. |
| Architecture of traffic control systems using cloud computing [23]. | 2010 | Traffic control system model utilizing cloud computing is proposed | can aid in the resolution of motor transport-related issues such as pollution, gridlock, auto theft, and road safety. | The use of cloud computing makes the system insecure. |
| An intelligent traffic and vehicle monitoring system using internet of things architecture [24]. | 2016 | Developed a model that uses an IoT board, microcontroller board, and GPS module. | vehicle spotting, and VIP and emergency vehicle clearance. | lack of emphasis on traffic signal control. |
| Real-time smart traffic management system for smart cities by using Internet of Things and big data [25]. | 2016 | Uses IoT to collect and process traffic data is presented | A mobile application is created as a user interface to investigate the density of traffic. | Emergency vehicles were not given clearance. |
| FC-PA: fog computing-based pseudonym authentication scheme in 5G-enabled vehicular networks [26]. | 2023 | Presents a fog computing-based pseudonym authentication (FC-PA) system for reducing performance overhead in 5G-enabled vehicle networks. | efficiency and security. | Higher communication and computing costs. |
| Chebyshev polynomial-based fog computing scheme supporting pseudonym revocation for 5G-enabled vehicular networks [27]. | 2023 | a fog computing strategy for 5G-enabled automotive networks that is based on the Chebyshev polynomial. | Improved performance and cost effectiveness. | Not concerned about privacy or security. |
| COVID-19 vehicle based on an efficient mutual authentication scheme for | 2022 | Presents a COVID-19 vehicle for 5G-enabled vehicular fog computing | More effective in terms of transmission and computing. | Inaccurate and higher cost. |

| 5G-enabled vehicular fog computing [28]. | | based on an effective mutual authentication system. | | |
|---|---|---|---|---|
| ANAA-Fog: A Novel Anonymous Authentication Scheme for 5G-Enabled Vehicular Fog Computing [29]. | 2023 | Propose ANAA-Fog, an anonymous authentication system for 5G-enabled vehicle fog computing. | Unlink ability, traceability, and conditional privacy preservation. | Higher costs for connectivity and computing performance. |
| Efficient authentication scheme for 5G-enabled vehicular networks using fog computing [30]. | 2023 | A fog server is employed in the proposed FC-CPPA approach to create a set of public anonymity. | Criteria for privacy and security. | Privacy and security objectives were not taken into account. |

## IV. ITCMS STRUCTURE

Our research investigates the proposed Intelligent Transportation and Congestion Management System (ITCMS), which aims to effectively control traffic and alleviate congestion. To showcase the optimization of traffic flow, we focus on a transportation-related example that demonstrates the simultaneous operations of fog nodes. The ITCMS is meticulously designed and implemented using iFogSim, incorporating various essential modules, including:

1. **The Fog Device Module:** This module is responsible for creating all fog devices and defining their hardware properties. These properties include device ID, MIPS (million instructions per second), RAM, uplink bandwidth, downlink bandwidth, level, rate per MIPS (cost rate per MIPS used), busy power (amperage rate per MIPS used), and the power consumption when the fog node is idle.
2. **The Sensor Module:** This module is used to create the required IoT sensors. The sensor can be connected to a router, fog node, or proxy via the gateway device. The setup link latency represents the time required to establish a connection between the sensor and the fog device. In the proposed ITCMS, a smart camera with two modules is utilized:
   - **Picture-capture module:** Integrated into the smart camera, this module captures images after a five-second delay, which are then transferred to the fog node.
   - **Slot detector module:** This module identifies empty traffic light slots.
3. **The Actuator Module**: This module generates objects that display output information. In the ITCMS scenario, LEDs serve as actuators to visually indicate the status of vacant traffic light slots (red LED, green LED, and yellow LED). Actuators need to be connected to a gateway device, through which data is transmitted. Therefore, when configuring actuators in iFogSim, it is necessary to specify the gateway device and the latency of the link.

The ITCMS comprises the following components:
- A cloud server
- Fog nodes
- Smart camera
- Three LED display screens

The smart camera is positioned near the traffic lights to capture images of vehicles, which are then transmitted to the fog node. On the fog node, an image processing method is implemented to identify vacant slots near the traffic lights. Once the vacant slots are detected, the relevant information is updated on the LED screens. The data is temporarily stored in the fog node before being transmitted to the cloud server for permanent storage. This enables drivers to promptly identify available spots upon reaching the traffic lights and move their vehicles to the designated location.

The information displayed on the three LED screens is refreshed every five seconds. To facilitate communication between the fog node and the cloud server, a proxy server is employed.

By employing this comprehensive system architecture, our proposed ITCMS aims to provide an intelligent and efficient solution for traffic congestion management.

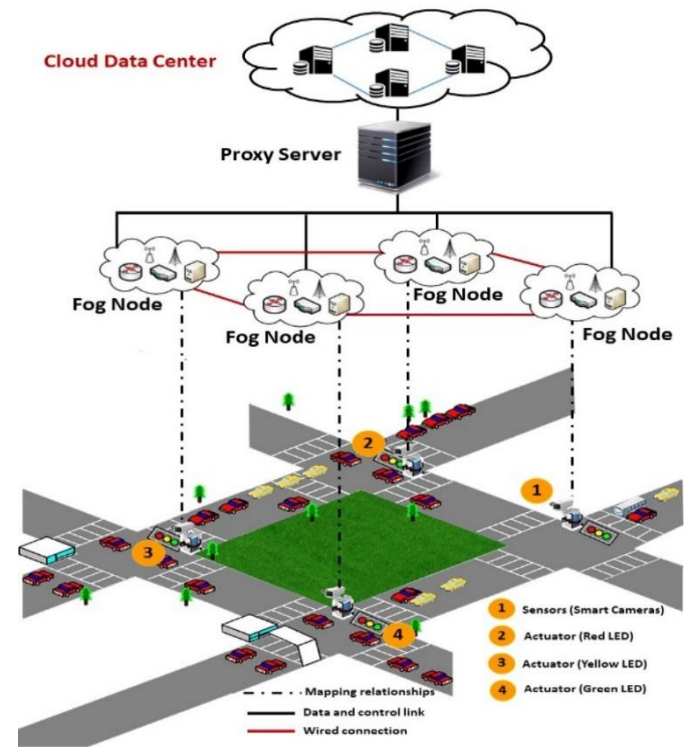

Fig. 2 ITCMS based Intelligent traffic congestion mitigation.

## V. THE PROTOTYPE OF THE ITCMS IMPLEMENTATION

The implementation of the proposed Intelligent Transportation and Congestion Management System (ITCMS) prototype takes place in an environment consisting of four roads: x, y, z, and w. In this configuration, the presence of multiple cars simultaneously results in traffic congestion. This setup is visually depicted in Fig. 3. Each fog node is equipped with a camera and three LEDs: Yellow, Green, and Red.

As a car moves from the source point (R1) to its destination (R2), the corresponding LED in the fog node illuminates green, indicating the availability of a clear path. Meanwhile, the LEDs in the other roads display red signals, indicating that those roads are congested. A detailed description of the proposed system can be found in Algorithm 1, providing a comprehensive understanding of its functioning and operation.

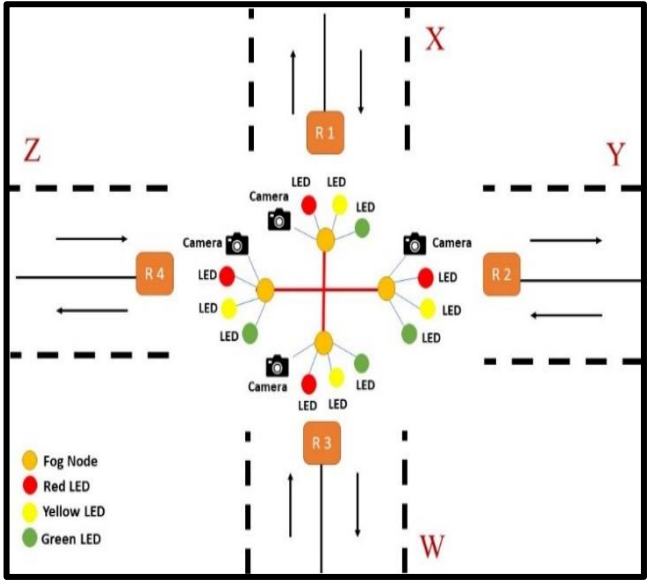

Fig. 3 The four fog nodes for four roads

**Algorithm 1: Path Function**
**Input: R1, R2, R3, R4 and Cameras**
**Output: Led1(Red), Led2(Yellow), Led3(Green)**
**Function: path (S, D)**
**S: Source**
**D: Destination**
1: *Path (X, Y)*
2: *{*
3: **Let led of R**1 = R2 = R3 = R4 = *Led3(Green)*,
4: if (X ⟶ Y)
5: Where X(car) in R1, Y(location) in R2
6: Then
7: *Led of R1 = Led3(Green), led of R2=R3=R4 =Led1(Red)*
8: **else if (X ⟶ W)**
9: Where X(car) in R1, W(location) in R3
10: Then
11: *Led of R1 = Led3(Green), led of R2=R3=R4 =Led1(Red)*
12: **else if (X ⟶ Z)**
13: Where X(car) in R1, Z(location) in R4
14: Then
15: *Led of R1 = Led3(Green), led of R2=R3=R4 =Led1(Red)*
16: **end if**
17: if (Y ⟶ X)
18: Where Y(car) in R2, X(location) in R1
19: Then
20: *Led of R2 = Led3(Green), led of R1=R3=R4 =Led1(Red)*
21: **else if (Y ⟶ W)**
22: Where Y(car) in R2, W(location) in R3
23: Then
24: *Led of R2 = Led3(Green), led of R1=R3=R4 =Led1(Red)*
25: **else if (Y ⟶ Z)**
26: Where Y(car) in R2, Z(location) in R4
27: Then
28: *Led of R2 = Led3(Green), led of R1=R3=R4 =Led1(Red)*
29: **end if**
30: if (W ⟶ X)
31: Where W(car) in R3, X(location) in R1
32: Then
33: *Led of R3 = Led3(Green), led of R1=R2=R4 =Led1(Red)*
34: **else if (W ⟶ Y)**
35: Where W(car) in R3, Y(location) in R2
36: Then
37: *Led of R3 = Led3(Green), led of R1=R2=R4 =Led1(Red)*
38: **else if (W ⟶ Z)**
39: Where W(car) in R3, Z(location) in R4
40: Then
41: *Led of R3 = Led3(Green), led of R1=R2=R4 =Led1(Red)*
42: **end if**
43: if (Z ⟶ X)
44: Where Z(car) in R4, X(location) in R1
45: Then
46: *Led of R4 = Led3(Green), led of R1=R2=R3 =Led1(Red)*
47: **else if (Z ⟶ Y)**
48: Where Z(car) in R4, Y(location) in R2
49: Then
50: *Led of R4 = Led3(Green), led of R1=R2=R3 =Led1(Red)*
51: **else if (Z ⟶ W)**
52: Where Z(car) in R4, W(location) in R3
53: Then
54: *Led of R4 = Led3(Green), led of R1=R2=R3 =Led1(Red)*
55: **end if**

## VI. SIMULATION RESULTS

To alleviate traffic congestion, this paper presents the application of ITCMS to Almohafza Street in Mansoura, a city in Egypt with a population of approximately 6 million people. The experiments were carried out using the NetBeans IDE version 8.2 software on a DELL Latitude E6540 laptop. The iFogSim simulator, an open-source Java-based simulator developed by the Cloud Computing and Distributed Systems (CLOUDS) Laboratory at the University of Melbourne [31], was used for the simulations.

The iFogSim simulation is built on the fundamental framework of CloudSim, which is widely recognized as one of the most popular simulators for simulating cloud computing environments. iFogSim extends the abstraction of core CloudSim classes to enable the simulation of a customized fog computing environment, encompassing a multitude of IoT devices and fog nodes such as sensors and actuators.

The evaluation and analysis of the performance of the ITCMS in this paper rely on four crucial parameters: CPU usage, heap memory usage, throughput, and total average delay. These parameters play a vital role in assessing the system's performance and determining its effectiveness in alleviating traffic congestion. Furthermore, a comparative analysis has been conducted, comparing the ITCMS with IOV and STL, using two key parameters: throughput and total average delay. This comparison sheds light on the strengths and weaknesses of each system. The ITCMS itself depends on several parameters, including latency, traffic efficiency, average traffic flow rate, energy saving, and waiting time. The presented parameters in this paper are specifically designed to accurately measure and analyze the performance of the ITCMS, facilitating the identification of areas for improvement. By scrutinizing these parameters, it becomes feasible to evaluate the system's efficacy in reducing traffic congestion and enhancing traffic flow in densely populated cities:

1. Cloud node

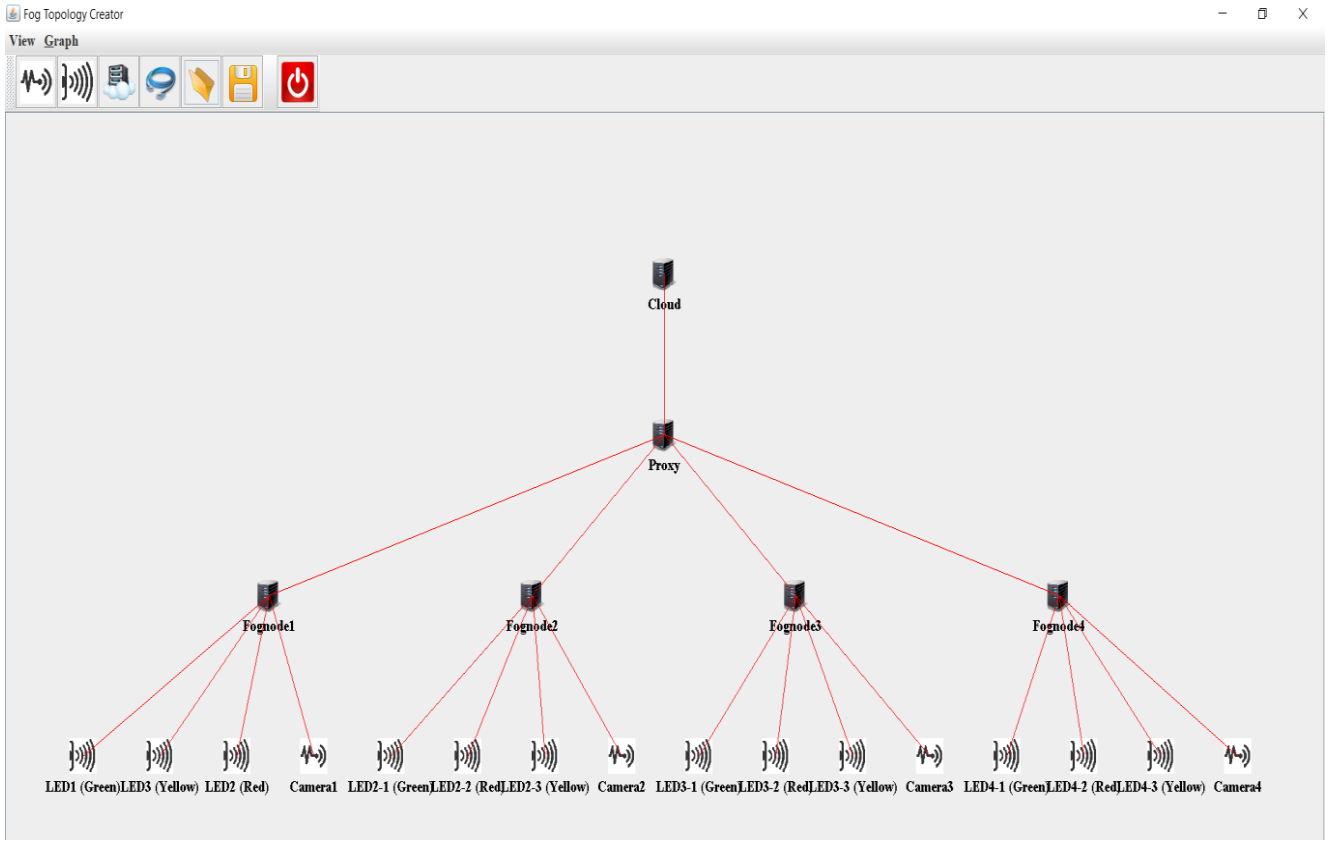

Fig. 4 The Simulated ITCMS for four roads

| Name | Cloud |
|---|---|
| Level | 0 |
| Uplink BW | 1000 |
| Downlink BW | 1200 |
| MIPS | 500 |
| RAM | 45000 |
| Rate/MIPS | 1000 |

2. Proxy node

| Name | Proxy |
|---|---|
| Level | 1 |
| Uplink BW | 1000 |
| Downlink BW | 1100 |
| MIPS | 4000 |
| RAM | 4500 |
| Rate/MIPS | 500 |

3. Fog node

| Name | Fog node |
|---|---|
| Level | 2 |
| Uplink BW | 800 |
| Downlink BW | 1000 |
| MIPS | 1000 |
| RAM | 3000 |
| Rate/MIPS | 400 |

4. LED or Camera node

| Name | LED or Camera |
|---|---|
| TYPE | SENSORS |
| Distribution type | Uniform |
| MIN | 20 |
| Max | 100 |

Cloud ⟶ Proxy, Latancy:200
Proxy ⟶ fog nodes, Latancy:100 Fog nodes ⟶ camera or led, Latancy:50

To simulate this scenario in iFogSim, it is necessary to generate a new class within the org.fog.test.Perceval package as depicted in Fig. 4. The FogDevice class facilitates the creation of fog nodes with different configurations through the utilization of a constructor. The provided code snippet below can be employed to generate heterogeneous fog devices.

Fig. 5 illustrates the CPU usage during the simulation, which lasted less than two minutes. The simulation began at 2:35:00 PM (Egypt Time), and the CPU usage started at 0% and gradually increased to 70% by 2:35:30 PM (Egypt Time). It then gradually decreased to 20% by 2:35:45 PM (Egypt Time) and ultimately reached 0% by 2:35:50 PM (Egypt Time).

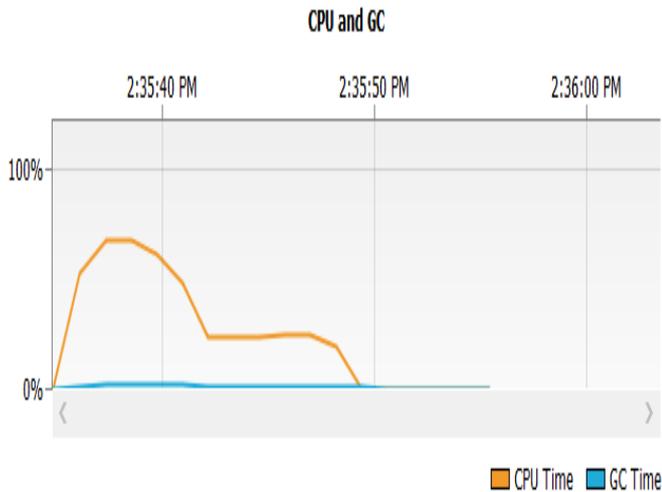

Fig.5 The CPU usage

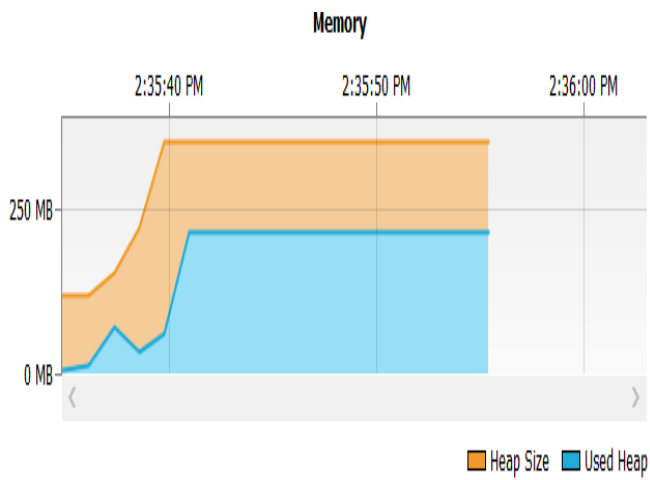

Fig. 6 The Heap Memory Usage

Fig. 6 shows the heap memory usage during the simulation. The used heap initially starts at 0 MB and gradually increases until it reaches 10 MB at 2:35:30 PM (Egypt Time). The used heap continues to increase until it reaches approximately 250 MB at 2:35:45 PM (Egypt Time). Finally, the used heap stabilizes at around 240 MB at the end of the simulation.

To verify the results of this study, eight fog nodes were deployed at eight roads, as shown in Fig. 7. Each road has a fog node with one camera and three LEDs. When a car travels from Road R1 (source) to Road R2 (destination), the LED in the fog node for Road R2 turns green, while the LEDs in the fog nodes for the other roads turn red.

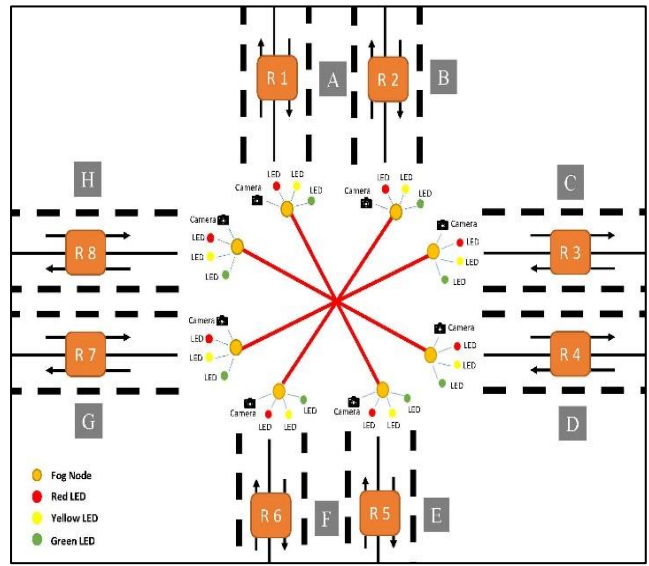

Fig. 7 The eight fog nodes for eight roads.

A. The CPU usage

The CPU usage metric provides valuable insights into the percentage of processing power utilized for data processing and program execution on a computer, server, or network device at any given time. This metric plays a crucial role in maintaining optimal performance and ensuring efficient system operation.

In essence, the CPU usage metric offers real-time information regarding the current utilization of processing power. This data enables the identification of potential bottlenecks and facilitates the implementation of corrective measures to enhance performance.

During the simulation depicted in Fig. 8, which lasted less than two minutes, the CPU usage was observed. The simulation commenced at 5:24:30 PM (Egypt Time) with an initial CPU usage of 0%. Subsequently, it gradually increased to 78% by 5:24:20 PM (Egypt Time). Following that, the CPU usage gradually decreased to 50% by 5:24:28 PM (Egypt Time) and eventually reached 0% by 5:24:30 PM (Egypt Time).

The spike in CPU usage observed at the beginning of the simulation can be attributed to the initialization of the ITCMS system. During this phase, the system loads data, initializes algorithms, and establishes connections with the fog nodes. Once the system is fully initialized, the CPU usage decreases as the system enters a stable state.

The gradual decrease in CPU usage throughout the simulation can be attributed to the effective utilization of resources by the ITCMS system. The system leverages a fog computing architecture to distribute the processing load across multiple devices. This approach enables the system to handle large data volumes without significantly impacting CPU usage.

The final decrease in CPU usage to 0% occurs upon the termination of the simulation. At this point, the ITCMS system releases all resources and returns to a low-power state.

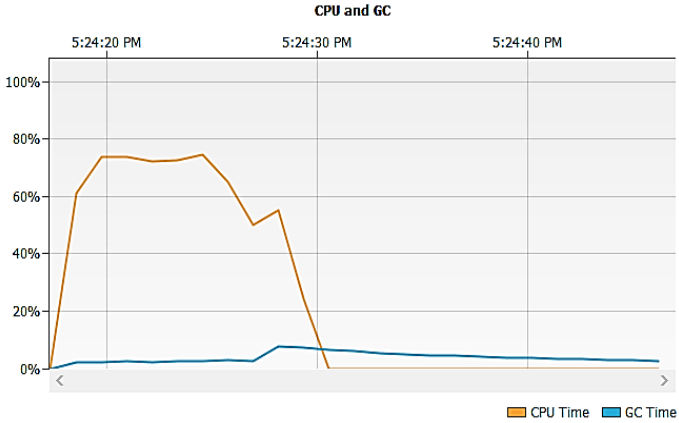

Fig. 8 The CPU usage

B. The Heap Memory Usage

The Java heap is a dedicated memory space specifically designed to store objects instantiated by applications running on the JVM. When the JVM is launched, a certain amount of memory is allocated for the heap, and any objects created can be shared among threads as long as the program is running. This memory space plays a critical role as it enables dynamic memory allocation, allowing objects to be created and destroyed in real-time as needed by the program.

The heap space is an integral component of the JVM runtime environment, and its efficient utilization is essential for ensuring optimal performance and minimizing memory-related issues such as memory leaks or out-of-memory errors. In essence, the Java heap provides a dedicated memory area for storing objects created by applications running on the JVM, facilitating efficient memory management and dynamic memory allocation.

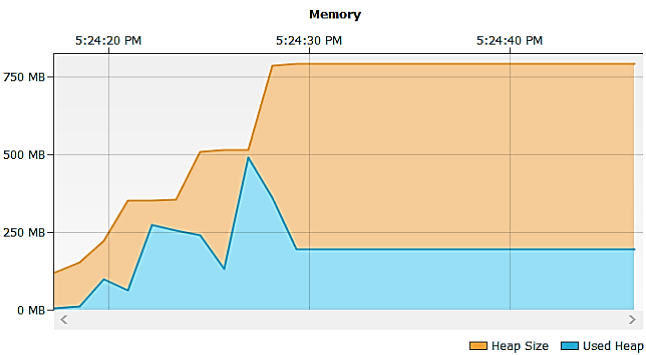

Fig. 9 The Heap Memory Usage

In Fig. 9, the Heap Memory Usage during the simulation is depicted. The Used Heap initially starts at 0 MB and gradually increases until it reaches 10 MB by 5:24:19 PM (Egypt Time). Subsequently, the Used Heap continues to grow, reaching approximately 500 MB by 5:24:28 PM (Egypt Time). Finally, at the end of the simulation, the Used Heap stabilizes at around 240 MB.

C. The Throughput

The system's throughput represents the rate at which cars pass through the intersection per second. To compare the proposed ITCMS system with previous traffic light management scenarios, the simulation utilized the IoV [32] and STL [33] systems.

In the IoV system, it is assumed that cars enter the road every 2.5 seconds, with a car size of 4.5 meters and a gap of 0.5 meters between them. This implies that the maximum number of cars on a single road is 80, and for two roads, it is 160.

The STL system calculates the overall time for a single cycle using basic mathematical computations. STL estimates that the traffic light would remain open for 30 seconds, with four roads converging. If two cars arrive every 15 seconds and three cars depart every 6 seconds at green lights on each road, the total time for a single cycle is (30*4=120 seconds). The average number of cars arriving in the signal cycle is (90/15)*2=12. The number of cars exiting during the 30-second green light period is (30/6)*3=15 cars. The only modification in this system is extending the duration of the green traffic light by 16 seconds to prevent congestion.

Throughput was measured for each system using relevant formulas and algorithms. The results for each system demonstrate the relationship between time and the number of cars crossing the intersection.

Fig. 10 illustrates the number of cars passing through the road intersection per second using the proposed ITCMS system. The ITCMS system achieves benchmark throughput values, as mentioned in [4]. As depicted in the figure, IoV and STL exhibit lower throughput compared to the ITCMS system. This is attributed to the ITCMS system's more accurate prediction of the optimal waiting time for each road, enabling a more efficient utilization of the green light duration.

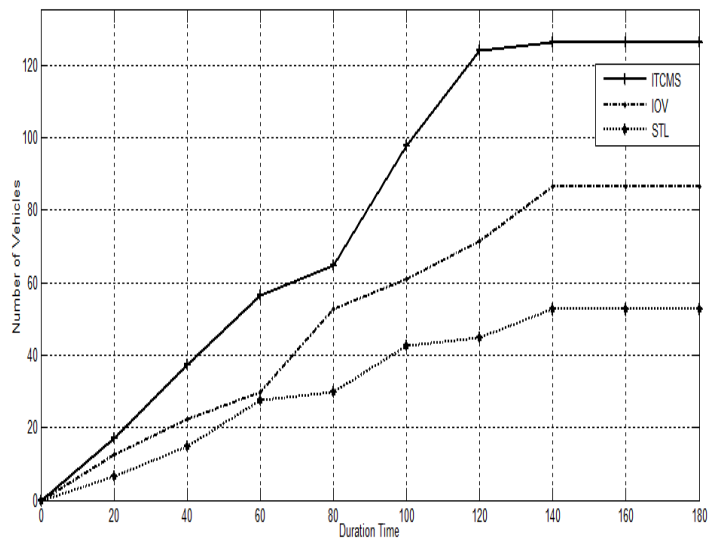

Fig. 10 The Throughput of the traffic intersection

The accuracy of the ITCMS system has been demonstrated because the number of crossing cars exceeds the number of waiting cars, as indicated in the throughput. This means that the ITCMS system is able to effectively reduce traffic congestion.

D. The total average delay

Addressing delays in traffic management systems poses a challenging and delicate task that necessitates a dynamic and speedy network, along with lightweight algorithms. While various strategies and procedures exist to tackle traffic congestion, only a few of them consider guarantees for overall average delay. The ITCMS System asserts its ability to achieve guaranteed latency by receiving real-time stream data and utilizing fog computing to predict the optimal waiting time. This fog-based algorithm eliminates the need to waste time gathering and transmitting data to remote servers, resulting in faster decision-making processes.

The red traffic duration is set to enable cars from each road to cross the intersection. As depicted in Fig. 11, even with a high number of cars in the simulation process, the ITCMS System exhibits acceptable delay times. This is attributed to the appropriate selection of a 5-second yellow light duration and the system's mobility. The total delay experienced by the ITCMS System is 30% less than the required delay of the IoV system and 60% less than that of the STL system. Consequently, the proposed ITCMS System claims to outperform IoV and STL in terms of reducing the average delay per car.

The findings indicate that the proposed ITCMS system accurately computes the optimal waiting time for each road, effectively extending the green light duration for a specific road as the number of cars increases.

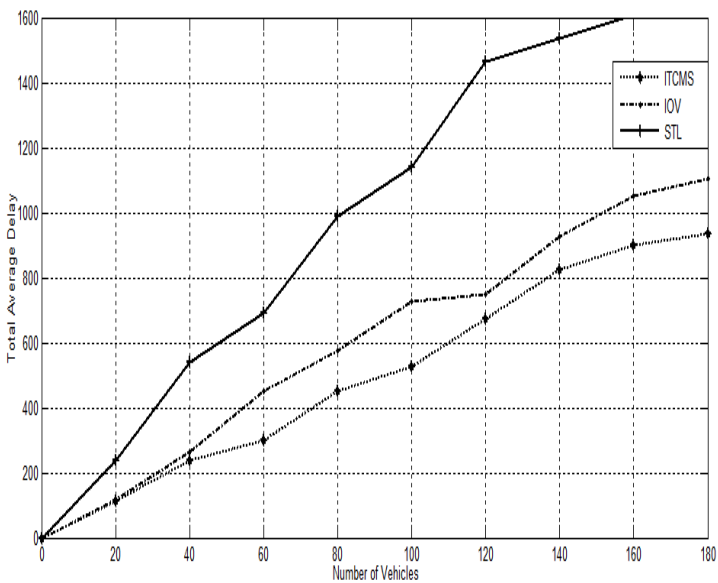

Fig. 11 The total average delay

In summary, the results of our proposed system were compared with those of previous studies in the literature, including the study conducted by Mohammed, T. S. et al. [34]. Through this comparison, it was observed that our system significantly outperformed the performance of the previous study by a 70% margin. Specifically, while Mohammed, T. S. et al. reported a latency of 11 seconds, our study demonstrated a much-improved latency of only 4 seconds. These findings highlight the superior performance of our proposed system and the significant advancements it offers in comparison to previous research in this field.

To further validate the findings of this study, we conducted simulations using three different fog node deployments: 4, 8, and 14 fog nodes. Each fog node was strategically positioned on a single road, as summarized in Table 2. Each road was equipped with a fog node, one camera, and three LEDs. Table 2 presents the simulation results for four key metrics: (a) execution time (ET), (b) application loop delay (ALD), (c) camera transmission time (CTT), and (d) total traffic flow (TTFU).

Table 2. Simulation results.

| NoFN | ET | ALD | CTT | TTFU |
|---|---|---|---|---|
| 4 | 4,538 | 49.67 | 5 | 3184 |
| 8 | 12,102 | 51.10 | 5 | 12736 |
| 14 | 48,481 | 53.25 | 5 | 39004 |

Fig. 12 illustrates the relationship between execution time and the number of fog nodes, showcasing three distinct scenarios with 4, 8, and 14 fog nodes.

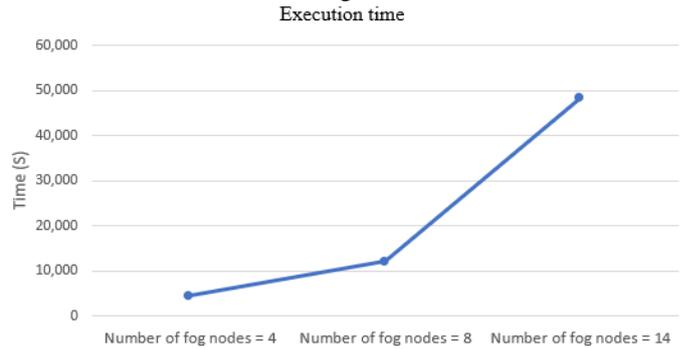

Fig. 12 The execution time.

Fig. 13 depicts the application loop delay for three different fog node configurations, encompassing 4, 8, and 14 fog nodes.

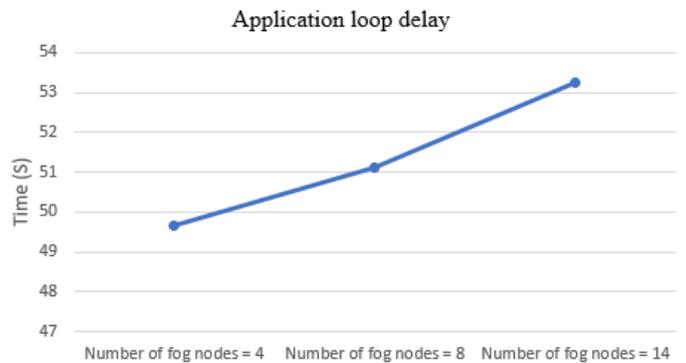

Fig. 13 The application loop delay

Fig. 14 illustrates the camera transmission time for three distinct fog node deployments, comprising 4, 8, and 14 fog nodes.

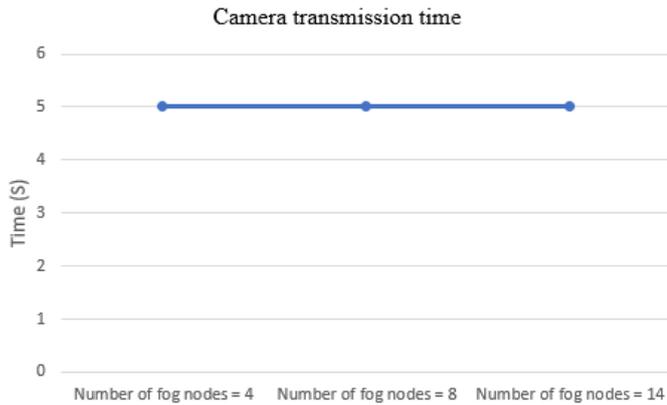
Fig. 14 Camera transmission time

Fig. 15 depicts the total traffic flow usage for three distinct fog node configurations, encompassing 4, 8, and 14 fog nodes.

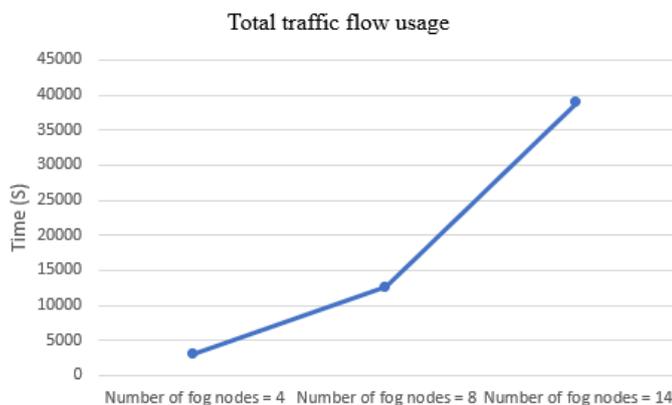
Fig. 15 Total traffic flow usage

## VII. CONCLUSION AND FUTURE WORKS

This paper presents a novel approach to alleviating traffic congestion by strategically deploying fog nodes at traffic intersections. These fog nodes are tasked with gathering, analyzing, and processing real-time traffic data. The findings demonstrate that congestion can be effectively mitigated by leveraging the average traffic flow rate between fog nodes and latency. To assess the performance of the proposed ITCMS system, four key metrics are employed: CPU usage, heap memory usage, throughput, and total average end-to-end delay. A comparative analysis of the ITCMS, IOV, and STL systems is also conducted using throughput and total average end-to-end delay metrics to identify the strengths and limitations of each system. Future research directions include implementing the proposed system in large and densely populated cities like Cairo to further evaluate its effectiveness in reducing traffic congestion.

## REFERENCES

[1] R. K. Naha *et al.*, "Fog Computing: Survey of Trends, Architectures, Requirements, and Research Directions," *IEEE Access,* vol. 6, pp. 47980-48009, 2018, doi: 10.1109/access.2018.2866491.

[2] A. De Mauro, M. Greco, and M. Grimaldi, "A formal definition of Big Data based on its essential features," *Library Review,* vol. 65, no. 3, pp. 122-135, 2016/04/04 2016, doi: 10.1108/lr-06-2015-0061.

[3] R. Matos, J. Araujo, D. Oliveira, P. Maciel, and K. Trivedi, "Sensitivity analysis of a hierarchical model of mobile cloud computing," *Simulation Modelling Practice and Theory,* vol. 50, pp. 151-164, 2015/01 2015, doi: 10.1016/j.simpat.2014.04.003.

[4] P. Pereira, J. Araujo, and P. Maciel, "A Hybrid Mechanism of Horizontal Auto-scaling Based on Thresholds and Time Series," presented at the 2019 IEEE International Conference on Systems, Man and Cybernetics (SMC), 2019/10, 2019. [Online]. Available: http://dx.doi.org/10.1109/smc.2019.8914522.

[5] A. El Shenawy, K. Mohamed, and H. Harb, "HDec-POSMDPs MRS Exploration and Fire Searching Based on IoT Cloud Robotics," *International Journal of Automation and Computing,* vol. 17, no. 3, pp. 364-377, 2019/07/15 2019, doi: 10.1007/s11633-019-1187-6.

[6] M. Laroui, B. Nour, H. Moungla, M. A. Cherif, H. Afifi, and M. Guizani, "Edge and fog computing for IoT: A survey on current research activities & future directions," *Computer Communications,* vol. 180, pp. 210-231, 2021/12 2021, doi: 10.1016/j.comcom.2021.09.003.

[7] J. P. Shim, R. Sharda, A. M. French, R. A. Syler, and K. P. Patten, "The Internet of Things: Multi-faceted Research Perspectives," *Communications of the Association for Information Systems,* pp. 511-536, 2020, doi: 10.17705/1cais.04621.

[8] C.-H. Hong and B. Varghese, "Resource Management in Fog/Edge Computing," *ACM Computing Surveys,* vol. 52, no. 5, pp. 1-37, 2019/09/13 2019, doi: 10.1145/3326066.

[9] M. Al Yami and D. Schaefer, "Fog Computing as a Complementary Approach to Cloud Computing," presented at the 2019 International Conference on Computer and Information Sciences (ICCIS), 2019/04, 2019. [Online]. Available: http://dx.doi.org/10.1109/iccisci.2019.8716402.

[10] A. Sunyaev, "Fog and Edge Computing," in *Internet Computing*, ed: Springer International Publishing, 2020, pp. 237-264.

[11] T. A. Nguyen, D. Min, and E. Choi, "A Hierarchical Modeling and Analysis Framework for Availability and Security Quantification of IoT Infrastructures," *Electronics,* vol. 9, no. 1, p. 155, 2020/01/14 2020, doi: 10.3390/electronics9010155.

[12] P. Pereira, J. Araujo, M. Torquato, J. Dantas, C. Melo, and P. Maciel, "Stochastic performance model for web server capacity planning in fog computing," *The Journal of Supercomputing,* vol. 76, no. 12, pp. 9533-9557, 2020/02/28 2020, doi: 10.1007/s11227-020-03218-w.

[13] R. Deng, R. Lu, C. Lai, T. H. Luan, and H. Liang, "Optimal Workload Allocation in Fog-Cloud Computing Towards Balanced Delay and Power Consumption," *IEEE Internet of Things Journal,* pp. 1-1, 2016, doi: 10.1109/jiot.2016.2565516.

[14] S. Kunal, A. Saha, and R. Amin, "An overview of cloud-fog computing: Architectures, applications with security challenges," *Security and Privacy,* vol. 2, no. 4, 2019/05/16 2019, doi: 10.1002/spy2.72.

[15] A. A. Mutlag, M. K. Abd Ghani, N. Arunkumar, M. A. Mohammed, and O. Mohd, "Enabling technologies for fog computing in healthcare IoT systems," *Future Generation Computer Systems,* vol. 90, pp. 62-78, 2019/01 2019, doi: 10.1016/j.future.2018.07.049.

[16] Y. Alsaawy, A. Alkhodre, A. Abi Sen, A. Alshanqiti, W.

A. Bhat, and N. M. Bahbouh, "A Comprehensive and Effective Framework for Traffic Congestion Problem Based on the Integration of IoT and Data Analytics," *Applied Sciences,* vol. 12, no. 4, p. 2043, 2022/02/16 2022, doi: 10.3390/app12042043.

[17] Qin H, Zhang H, Intelligent traffic light under fog computing platform in data control of real-time traffic flow, J Supercomput 77(5):4461–4483, 2021.

[18] Maria Viorela Muntean, Multi-Agent System for Intelligent Urban Traffic Management Using Wireless Sensor Networks Data, Sensors 2022, 22(1), 208; https://doi.org/10.3390/s22010208 , 2022.

[19] Golhar, Yogesh, and Manali Kshirsagar, Efficient Strategies to Manage Road Traffic Using Big Data Analytics, ICT Analysis and Applications. Springer, Singapore, 2022, p.47-55.

[20] M. Gupta, D. Kumar, and M. Kumar, "IOT-Based Smart Traffic Light System for Smart Cities," in *Algorithms for Intelligent Systems*, ed: Springer Singapore, 2021, pp. 579-585.

[21] M. M. Rathore, A. Paul, S. Rho, M. Khan, S. Vimal, and S. A. Shah, "Smart traffic control: Identifying driving-violations using fog devices with vehicular cameras in smart cities," *Sustainable Cities and Society,* vol. 71, p. 102986, 2021/08 2021, doi: 10.1016/j.scs.2021.102986.

[22] J. Cunha, C. Cardeira, and R. Melício, "Traffic Lights Control Prototype Using Wireless Technologies," *Renewable Energy and Power Quality Journal,* pp. 1031-1036, 2016/05 2016, doi: 10.24084/repqj14.562.

[23] V. V. Brizgalov, V. Chukhantsev, and E. Fedorkin, "Architecture of traffic control systems using cloud computing," presented at the 2010 11th International Conference and Seminar on Micro/Nanotechnologies and Electron Devices, 2010/06, 2010. [Online]. Available: http://dx.doi.org/10.1109/edm.2010.5568826.

[24] B. K. Perumalla and M. S. Babu, "An intelligent traffic and vehicle monitoring system using internet of things architecture," *Int. J. Sci. Res,* vol. 5, no. 55, pp. 213-216, 2016.

[25] P. Rizwan, K. Suresh, and M. R. Babu, "Real-time smart traffic management system for smart cities by using Internet of Things and big data," presented at the 2016 International Conference on Emerging Technological Trends (ICETT), 2016/10, 2016. [Online]. Available: http://dx.doi.org/10.1109/icett.2016.7873660.

[26] Mohammed, B.A, et al. "FC-PA: fog computing-based pseudonym authentication scheme in 5G-enabled vehicular networks." IEEE Access 11 (2023): 18571-18581.

[27] Al-Mekhlafi, Z.G, et al. "Chebyshev polynomial-based fog computing scheme supporting pseudonym revocation for 5G-enabled vehicular networks." Electronics 12.4 (2023): 872.

[28] Al-Shareeda, M.A.& Manickam,S. "COVID-19 vehicle based on an efficient mutual authentication scheme for 5G-enabled vehicular fog computing." International journal of environmental research and public health 19.23 (2022): 15618.

[29] Mohammed, B.A, et al. "ANAA-Fog: A Novel Anonymous Authentication Scheme for 5G-Enabled Vehicular Fog Computing." Mathematics 11.6 (2023): 1446.

[30] Al-Mekhlafi, Z.G, et al. "Efficient authentication scheme for 5G-enabled vehicular networks using fog computing." Sensors 23.7 (2023): 3543.

[31] R. Mahmud and R. Buyya, "Modeling and Simulation of Fog and Edge Computing Environments Using iFogSim Toolkit," in *Fog and Edge Computing*, ed: John Wiley & Sons, Inc., 2019, pp. 433-465.

[32] S. A. Elsagheer Mohamed and K. A. AlShalfan, "Intelligent Traffic Management System Based on the Internet of Vehicles (IoV)," *Journal of Advanced Transportation,* vol. 2021, pp. 1-23, 2021/05/26 2021, doi: 10.1155/2021/4037533.

[33] A. Alharbi, G. Halikias, A. A. A. Sen, and M. Yamin, "A framework for dynamic smart traffic light management system," *International Journal of Information Technology,* vol. 13, no. 5, pp. 1769-1776, 2021/07/29 2021, doi: 10.1007/s41870-021-00755-2.

[34] T. S. Mohammed, O. F. Khan, A. S. Ibrahim, and R. Mamlook, "Fog computing-based model for mitigation of traffic congestion," *Int. J. Simul. Syst. Sci. Technol,* vol. 19, no. 3, pp. 5.1-5.7, 2018.

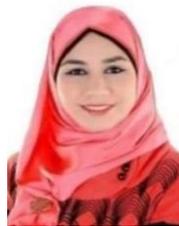

Alzahraa Elsayed received an MSc in Communications Engineering and Computers Engineering from the University of Al Azhar, Egypt (2018), where she is currently pursuing a Ph.D. in Communications Engineering from 2019 to 2022, her research interests include fog computing, cloud computing, and internet of things (IoT) technologies.

E-mail: alzahraa.salah@azhar.edu.eg

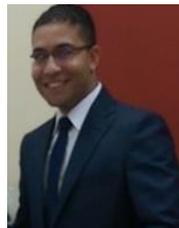

Khalil Mohamed received a Ph.D. in robotics and control engineering from Al-Azhar University, Egypt in 2019. He is currently an assistant professor at Systems and Computers Engineering Department, at Al-Azhar University, Egypt.

His research interests include AI, Machine learning, Deep learning, Reinforcement learning, Robotics, Control theory, Intelligent Control Systems, Automotive Control Systems, Robust Control, Stochastic Control, Motion and Navigation Control, Traffic and Transport Control, Predictive control, Optimal control, Mathematics, Optimization, Task assignment in multi-robot systems, Task decomposition.

E-mail: eng.khalil@azhar.edu.eg

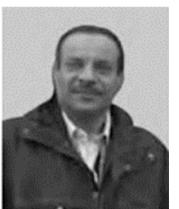

Hany Harb received a B.Sc. degree in computers and control engineering from the Faculty of Engineering, Ain Shams University, Egypt in 1978, and an M.Sc. degree in computers and systems engineering from the Faculty of Engineering, Al-Azhar University, Egypt in 1981. He also received a Ph.D. degree in computer science and an M.Sc. degree in operations research (MSOR) from the Institute of Technology (IIT), USA in 1986 and 1987, respectively. He is a professor of software engineering in the System Engineering Department, Faculty of Engineering, Al-Azhar University, Egypt. His research interests include artificial intelligence, cloud computing, and distributed systems.

E-mail: harbhany@yahoo.com